# Crab Waist collision scheme: a novel approach for particle colliders

**M Zobov on behalf of the DAΦNE Team**

LNF-INFN, Via Enrico Fermi 40, 00044 Frascati (Rome), Italy

E-mail: mikhail.zobov@lnf.infn.it

**Abstract**. A new concept of nonlinear focusing of colliding bunches, called Crab Waist (CW) collision scheme, has been proposed at LNF INFN. It has been successfully tested at the Italian lepton collider DAΦNE in operational conditions providing luminosity for two different experimental detectors, SIDDHARTA and KLOE-2. Considering a high efficiency of the scheme for increasing collision luminosity and its relative simplicity for implementation several new collider projects have been proposed and are under development at present. These are the SuperKEKB B-factory ready to start commissioning in 2016 in Japan, the SuperC-Tau factory proposed in Novosibirsk and entered in the short list of Russian mega-science projects, the new 100-km electron-positron Future Circular Collider (FCC-ee) under design study at CERN and some others. In this paper we describe the CW collision scheme, discuss its advantages and report principal results achieved at the electron-positron Φ-factory DAΦNE.

## 1. Introduction

The crab waist collision scheme (CW) was proposed [1, 2] and have been successfully tested at the Φ-factory DAΦNE [3]. At present this scheme is considered to be most attractive for the next generation lepton factories since it holds the promise of increasing the luminosity of the storage-ring colliders by 1-2 orders of magnitude beyond the current state-of-art, without any significant increase in beam current and without reducing the bunch length. Several new collider projects seek to exploit the potential of the crab waist collision scheme. In particular, physics and accelerator communities are discussing and developing new projects which make use of the CW collision scheme: SuperB-factory in Japan (SuperKEKB [4]), SuperC-Tau factory in Novosibirsk [5], two Higgs-factories FCC-ee at CERN [6] and CEPC in China [7].

The SuperKEKB is a natural upgrade of the very successful KEKB, Japanese B-factory at KEK (Tsukuba). The design luminosity goal of the project is $0.8 \times 10^{36}$ cm$^{-2}$s$^{-1}$, i.e. by a factor of 40 higher than the world record luminosity of $2.1 \times 10^{34}$ cm$^{-2}$s$^{-1}$ achieved at KEKB. The commissioning of SuperKEKB started in the very beginning of 2016 and the first beams have been already accumulated.

The Budker Institute of Nuclear Physics (Novosibirsk, Russia) is promoting the project of a new generation SuperC-Tau factory. The crab waist concept should allow reaching the project luminosity of $(1-2) \times 10^{35}$ cm$^{-2}$s$^{-1}$ that by more than 2 orders of magnitude higher than the luminosity of $0.8 \times 10^{33}$ cm$^{-2}$s$^{-1}$ presently achieved at the operating τ-charm factory BEPCII in Beijing [8]. SuperC-Tau has entered in the list of 6 most important Russian mega-science projects. In December 2015 the injection complex of the new collider has been successfully commissioned.

In 2014 CERN launched the Future Circular Collider (FCC) study aimed at the design of a 100-km proton-proton collider with the collision energy of 100 TeV. As an intermediate step, the electron-positron collider (FCC-ee) hosted in the same tunnel and covering the energy range between 90 GeV and 350 GeV is also under consideration and its intensive design study is ongoing. The CW scheme has been proposed [9] and recently chosen as the baseline option for the FCC-ee design. This scheme provides a substantially higher luminosity with respect to the traditional head-on collisions at low energies and approximately the same luminosity at higher energies (> 240 GeV) at much relaxed beam optics parameters. A CW lattice solution with twice higher vertical beta function at the interaction point, good dynamic aperture and energy acceptance and manageable photon energies has been found.

Presently China is considering to build a 54-km long Higgs factory CEPC (Circular Electron Positron Collider). Along with the head-on collision option the collider design team has started seriously evaluating a possibility to build a local double ring option with the Crab Waist interaction region [10].

In order to complete the picture, we should mention other collider projects that were considering an application of the CW scheme such as SuperB [11] and SuperTau-Charm [12] factories in Italy, a 500 GeV e+e- collider in a 233-km tunnel [13] in USA, and an upgrade option of the LHC based on collisions of very flat bunches [14, 15].

Since DAΦNE is the only collider exploiting the CW collisions, its operational experience is of great importance for all the new collider projects. In this paper we overview the CW concept and discuss principal results achieved during collider operation with SIDDHARTA and KLOE-2 experimental detectors.

## 2. Crab waist collision concept

The CW scheme can substantially increase collider luminosity since it combines several potentially advantageous ideas. Let us consider two bunches with the horizontal $\sigma_x$, vertical $\sigma_y$ and longitudinal $\sigma_z$ sizes colliding with a horizontal crossing angle $\theta$ (as shown in Fig. 1a).

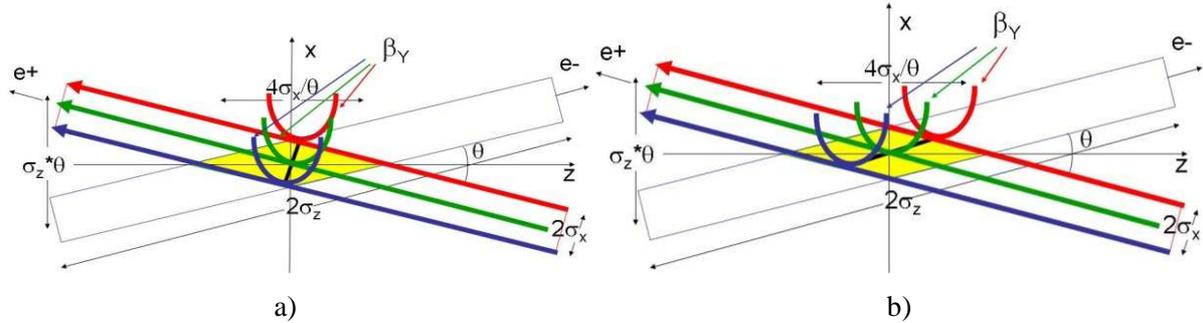

Figure 1. Crab Waist collision scheme: a) crab sextupoles OFF; b) crab sextupoles ON.

Then, the CW principle can be explained in the three basic steps. The **first one** is a large Piwinski angle $\Phi = (\sigma_z/\sigma_x)tg(\theta/2) \gg 1$. In the CW scheme the Piwinski angle is increased by decreasing the horizontal beam size and increasing the crossing angle. In this way we can gain in luminosity and the horizontal tune shift decreases; parasitic collisions (PC) become negligible since with higher crossing angle and smaller horizontal beam size the beam separation at the PC is larger in terms of $\sigma_x$. But the most important effect is that the overlap area of the colliding bunches is reduced, since it is proportional to $\sigma_x/\theta$ (see Fig. 1). As the **second step**, the vertical beta function $\beta_y$ can be made comparable to the overlap area size (i.e. much smaller than the bunch length):

$$\beta_y^* \approx \frac{2\sigma_x}{\theta} \cong \frac{\sigma_z}{\Phi} \ll \sigma_z$$

So, reducing $\beta_y^*$ at the IP gives us several advantages:
- Luminosity increase with the same bunch current.
- Possibility of the bunch current increase (if it is limited by $\xi_y$), thus farther increasing the luminosity.
- Suppression of the vertical synchrobetatron resonances [16].

Besides, there are additional advantages in such a collision scheme: there is no need in decreasing the bunch length to increase the luminosity as required in standard collision schemes. This will certainly help solving the problems of HOM heating, coherent synchrotron radiation of short bunches, excessive power consumption, etc.

However, implementation of these two steps introduces new beam-beam resonances which may strongly limit the maximum achievable tune shifts. At this point the crab waist transformation [2, 17] enters the game boosting the luminosity. This is the **third step**. As it is seen in Fig. 1b, the beta function waist of one beam is oriented along the central trajectory of the other one. In practice the CW vertical beta function rotation is provided by sextupole magnets placed on both sides of the IP in phase with the IP in the horizontal plane and at π/2 in the vertical one (as shown in Fig. 2).

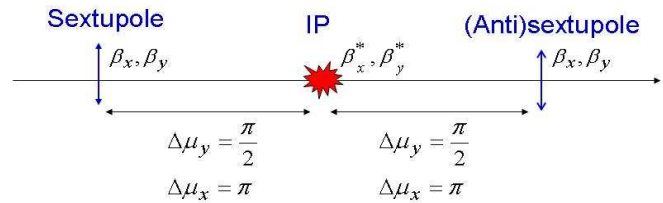

Figure 2. Crab sextupole locations with respect to the interaction point (IP).

The crab sextupole strength should satisfy the following condition depending on the crossing angle and the beta functions at the IP and the sextupole locations:

$$K = \frac{1}{\theta} \frac{1}{\beta_y^* \beta_y} \sqrt{\frac{\beta_x^*}{\beta_x}}$$

The crab waist transformation gives a small geometric luminosity gain due to the vertical beta function redistribution along the overlap area. It is estimated to be of the order of several percent. However, the dominating effect comes from the suppression of betatron (and synchrobetatron) resonances arising (in collisions without CW) due to the vertical motion modulation by the horizontal betatron oscillations [17]. Fig. 3 demonstrates the resonance suppression applying the frequency map analysis (FMA) for the beam-beam interaction in DAΦNE in CW collisions [18].

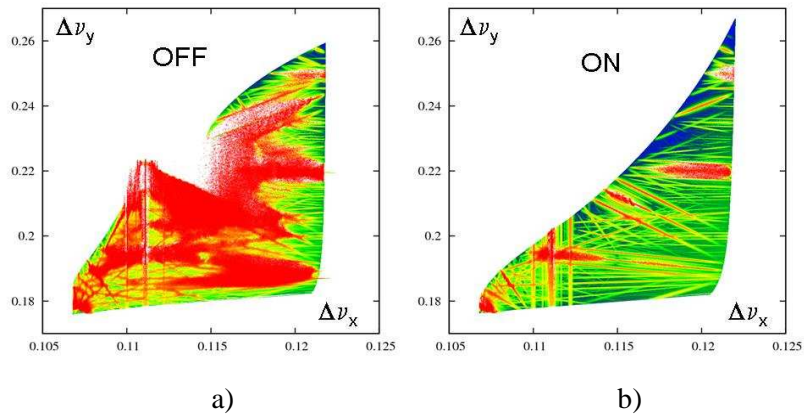

a)          b)

Figure 3. DAΦNE beam-beam footprint with crab sextupoles OFF (a) and ON (b) obtained by FMA.

## 3. DAΦNE operation with crab waist collision scheme

DAΦNE is an electron-positron collider working at the energy of the Φ resonance (1.02 GeV c.m.) aimed at providing a high rate of K mesons production [19, 20]. DAΦNE complex consists of two main rings and an injection system composed of a full energy linear accelerator, a damping/accumulator ring and transfer lines. After relevant collider modifications [21, 22] the crab waist collision scheme was successfully implemented and tested at DAΦNE providing luminosity increase by a factor of 3 for the SIDDHARTA experiment [3], in a good agreement with numerical simulations [23].

Figure 4 shows the effect of the crab sextupoles on the beam-beam blow up and the distribution tails in DAΦNE. The main results achieved during the experimental run in terms of peak, specific and integrated luminosity are listed in Table 1.

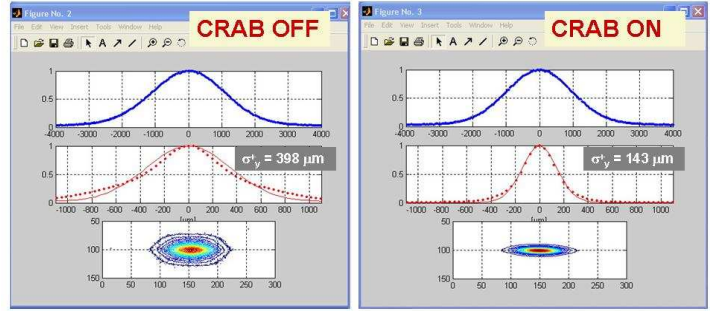

Figure 4. Transverse beam profiles with crab sextupoles OFF (left) and ON (right).

After completion of the SIDDHARTA experimental run and the collider consolidation program [24] the new, more complex, KLOE-2 experimental detector was installed at DAΦNE. Integrating the CW scheme with the large detector having a strongly perturbing solenoidal field posed new challenging issues concerning machine layout, beam acceptance and coupling correction. A new interaction region has been designed and has been equipped with a transverse betatron correction mechanism based on rotated quadrupoles and anti-solenoids independent for the two beams [25].

Table 1. Best luminosity performances during SIDDHARTA and KLOE-2 experimental runs

| Parameter | SIDDHARTA | KLOE-2 |
|---|---|---|
| Peak luminosity [$cm^{-2}s^{-1}$] | $4.53 \times 10^{32}$ | $2.13 \times 10^{32}$ |
| Electron beam current [A] | 1.52 | 1.13 |
| Positron beam current [A] | 1.00 | 0.88 |
| Number of bunches | 105 | 105 |
| Specific luminosity [$cm^{-2}s^{-1}mA^{-2}$/bunch] | $3.13 \times 10^{28}$ | $2.25 \times 10^{28}$ |
| Integrated luminosity [$pb^{-1}$/day] | 14.98 | 14.03 |

Despite the presence of the rotated quadrupoles, strong detector solenoid with nonlinear fringing fields and anti-solenoids between the crab sextupoles the effectiveness of the CW scheme has been proven both numerically [26] and experimentally [24]. As it is seen in Table 1, the presently achieved daily integrated luminosity is comparable to that obtained during the run with a simpler SIDDHARTA interaction region. However, the maximum peak luminosity is still by about a factor of 2 lower than in the previous SIDDHARTA run. Luminosity limiting factors have been identified and a work is in progress to eliminate them in order to fully exploit the CW potential. In particular, the following measures are foreseen to improve the DAΦNE performance: improving CW sextupoles alignment and their strength optimization; refining the transverse coupling correction; pushing the microwave instability threshold towards higher single bunch currents by using optics with a higher momentum compaction factor and a higher chromaticity; vacuum conditioning and scrubbing to diminish electron cloud effects; further feedback noise reduction etc.

## 4. Conclusions

Experimental measurements at DAΦNE have proven that the Crab Waist is a reliable and robust technique for the luminosity increase. It has been very successful in collisions with SIDDHARTA interaction region. The obtained results are in a very good agreement with numerical predictions and simulations. The work is now in progress to fully exploit the potential of the Crab Waist in collisions for the KLOE-2 experimental detector. At present several projects of large scale colliders based on the Crab Waist Scheme are under development.